# Measurements of polarization-dependent angle-resolved light scattering from individual microscopic samples using Fourier transform light scattering

**JaeHwang Jung[1,2], Jinhyung Kim[3], Min-Kyo Seo[3], and YongKeun Park[1,2,4,*]**

[1]*Department of Physics, Korea Advanced Institute of Science and Technology (KAIST), Daejeon 34141, Republic of Korea*
[2]*KAIST Institute for Health Science and Technology, KAIST, Daejeon 34141, Republic of Korea*
[3]*Department of Physics and Institute for the NanoCentury, Korea Advanced Institute of Science and Technology, Daejeon 305-701, South Korea*
[4]*Tomocube, Inc., Daejeon 34051, Republic of Korea.*
[*]*yk.park@kaist.ac.kr*

**Abstract:** We present a method to measure the vector-field light scattering of individual microscopic objects. The polarization-dependent optical field images are measured with quantitative phase imaging at the sample plane, and then numerically propagated to the far-field plane. This approach allows the two-dimensional polarization-dependent angle-resolved light scattered patterns from individual object to be obtained with high precision and sensitivity. Using this method, we present the measurements of the polarization-dependent light scattering of a liquid crystal droplet and individual silver nanowires over scattering angles of 50°. In addition, the spectroscopic extension of the polarization-dependent angle-resolved light scattering is demonstrated using wavelength-scanning illumination.

## 1. Introduction

Electromagnetic (EM) elastic scattering from microscopic objects is a physical phenomenon of fundamental importance. It has been utilized in various disciplines, such as material science [1, 2], soft matter physics [3, 4], biophysics [5], and tissue inspections [6]. Measurements of angle-resolved light scattering (ALS) patterns from micrometer-sized samples can be used to identify important properties of the objects, including the size, shape, and refractive index [7]. Conventionally, the ALS pattern has been measured using goniometer-based instruments [8] or microscopic imaging at back-focal planes [9, 10].

Despite its importance in various applications, measurements of polarization-sensitive ALS signals from individual objects have remained challenging, especially for microscopic individual objects with small scattering cross-sections, because of the following reasons. First, it is difficult to isolate a microscopic object without imaging the specific object at the microscopic resolution. Secondly, the polarization direction of an incident beam should be adjusted to the corresponding object. Thirdly, the power of the scattering signal from the object is small and exhibits extremely large dynamic ranges.

This measurement constraint is unfortunate because polarization-sensitive ALS signals from individual objects provide invaluable information about objects exhibiting birefringence [11, 12]. Despite the many challenges, there is strong motivation to access polarization-sensitive ALS signals from microscopic objects, especially of weakly scattering and transparent characteristics.

In this paper, we present the experimental measurements of polarization-dependent two-dimensional (2D) ALS signals from individual microscopic objects, using polarization-dependent quantitative phase imaging (QPI) [13, 14] and Fourier-transform light scattering (FTLS) techniques [15]. In the present method, hereafter referred to as polarization Fourier transform light scattering (pFTLS), the polarization-dependent optical field maps, consisting both the amplitude and phase images, of a sample are measured using the QPI, from which the polarization-dependent ALS signal is retrieved via numerical propagation. To verify the capability of the pFTLS technique, we measured the polarization-dependent scattered electric fields of a liquid crystal (LC) droplet and individual silver nanowires (NW). Moreover, the spectroscopic extension of the pFTLS method is demonstrated by using the wavelength-scanning illumination.

## 2. Method

### 2.1 Polarization-sensitive diffraction phase microscopy

To measure polarization-sensitive optical field images, we developed polarization-sensitive diffraction phase microscopy (DPM) [Fig. 1(a)]. DPM is a common-path QPI technique capable of measuring optical field images of a sample with high stability [16, 17]. To measure complete polarization-dependent optical field information, a polarizer and an analyzer were added and systematically controlled in front of a sample and an image sensor respectively [Fig. 1(a)].

In DPM, a spatially filtered plane wave impinges onto the sample. The diffracted light is collected and projected via an infinity-corrected imaging system to an imaging plane, which is then diffracted by a grating into several diffraction orders. Among them, the $0^{th}$ and $1^{st}$ diffraction orders are collected and imaged onto the image sensor. The $0^{th}$ diffraction beam is spatially filtered at Fourier plane so that it becomes a plane wave at the image plane. The sample (the $1^{st}$ diffraction order) and reference (the $0^{th}$ diffraction order) beams interfere with each other at the image plane, and the interferogram is recorded by the image sensor.

To measure the polarization-dependent optical fields, the polarizer and analyzer were placed before the sample and the camera plane, respectively [Fig. 1(b)]. Four complex optical fields $E_1$, $E_2$, $E_3$, and $E_4$ were measured under four different polarization conditions: (+45°, 0°), (+45°, +90°), (-45°, 0), and (-45°, +90°), where the angles in parentheses represent the direction of the polarizer and analyzer, respectively.

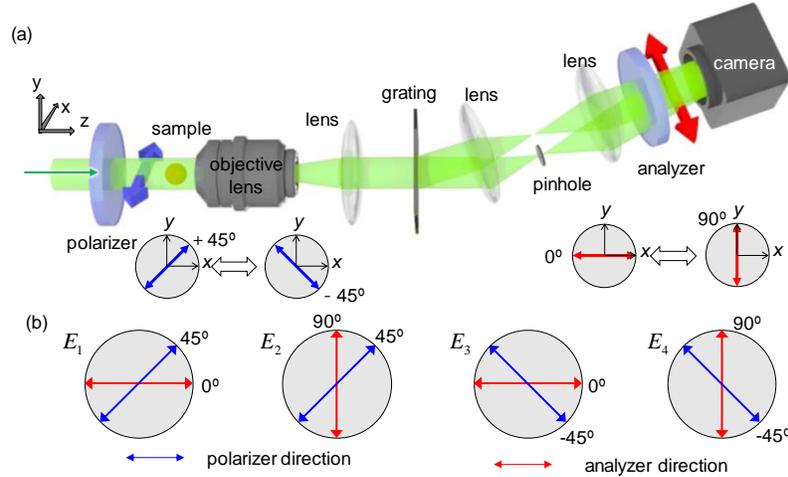

Fig. 1. Schematic of the experimental setup. (a) Polarization-dependent optical field images are measured using diffraction phase microscopy equipped with a polarizer and an analyzer. The orientations of the polarizer are switched between +45° and -45°. The orientations of the analyzer were switched between 0° and 90°. (b) Four configurations for the polarizer and analyzer to measure four different optical fields, $E_1$, $E_2$, $E_3$, and $E_4$.

For the LC droplet measurements, a He-Ne laser ($\lambda$ = 633 nm, 5 mW, Thorlabs Inc.) was used as an illumination source. Diffracted light from the sample was collected by a 100× oil immersion objective lens (UPLSAPO 100XO, 1.4 NA, Olympus). The interferograms were measured using a CMOS camera (FASTCAM 1024 PCI, Photron Inc.). In total, the measurements of $E_1$, $E_2$, $E_3$, and $E_4$ required 32 ms.

For the silver NW measurements, we built a wavelength-swept source based on a supercontinuum (SuperK COMPACT, NKT Photonics) for spectral demonstration. Utilizing a prism and a pinhole, the wavelength-swept source produced light with a narrow spectral bandwidth (1.2–6.0 nm as the wavelength increased from 460 to 660 nm) [18]. Diffracted light from the sample was collected via a 60× oil immersion objective lens (PLAPON 60XO, 1.42 NA, Olympus). The interferograms were measured by a sCMOS camera (Neo 5.5, Andor Technology Ltd.). For the spectroscopic demonstration, total acquisition time was ~30 s; four complex fields covering the 50 degrees of scattering angles were measured for each wavelength, and the measurements were repeated for 50 wavelengths. Further information about the optical system and the phase retrieval algorithm can be found in [18-20].

*2.2 Fourier transform light scattering*

FTLS is a numerical method for retrieving a far-field scattering pattern from QPI [15, 21]. From the optical field image of a sample measured at a focal plane $E(x, y)$, the optical field at a far-field plane $\tilde{E}(u,v)$ is obtained via Fraunhofer diffraction equation:

$$\tilde{E}(u,v) \propto \frac{n_m}{\lambda} \iint E(x,y) \exp\left[-2\pi i (ux+vy)\right] dx dy, \qquad (1)$$

where $u = n_m \sin\theta \cos\varphi / \lambda$ and $v = n_m \sin\theta \sin\varphi / \lambda$ are spatial frequencies along the $x$- and $y$-axis, respectively. $n_m$, $\lambda$, $\theta$, and $\varphi$ denote the refractive index of a medium, the wavelength of light in vacuum, the polar angle, and the azimuthal angle, respectively. Equation (1) has a mathematical form equivalent to 2D spatial Fourier transform.

Compared to conventional goniometer-based instruments, FTLS has several advantages in measuring the ALS patterns from micrometer-sized samples: (i) it can isolate individual microscopic objects with QPI; (ii) the dynamic range of ALS measurements is very high, which is determined by the pixel resolution, rather than the dynamic range, of the image sensor; (iii) the sample orientation alignments to the polarization direction of the probe beam is available. Due to these unique advantages, FLTS has been widely used in various applications, including tissue slices [22, 23], blood cells [24-26], colloidal clusters [27], and bacteria [28].

*2.3 The principle of pFLTS*

The polarization of the light interacting with a sample can be described by a Jones matrix. From measurements of the four fields, $E_1$, $E_2$, $E_3$, and $E_4$, the spatially resolved 2D Jones matrix can be reconstructed as:

$$\begin{bmatrix} J_{HH}(x,y) & J_{HV}(x,y) \\ J_{VH}(x,y) & J_{VV}(x,y) \end{bmatrix} = \frac{1}{2}\begin{bmatrix} E_1(x,y)+E_3(x,y) & E_1(x,y)-E_3(x,y) \\ E_2(x,y)+E_4(x,y) & E_2(x,y)-E_4(x,y) \end{bmatrix}. \qquad (2)$$

For a plane wave linearly polarized to an angle $\alpha$ with respect to the $x$-axis, the electric field at the sample plane can be expressed as:

$$\begin{bmatrix} E_H(x,y) \\ E_V(x,y) \end{bmatrix} = \begin{bmatrix} J_{HH}(x,y) & J_{HV}(x,y) \\ J_{VH}(x,y) & J_{VV}(x,y) \end{bmatrix} \begin{bmatrix} \cos\alpha \\ \sin\alpha \end{bmatrix}, \qquad (3)$$

where subscript $H$ and $V$ denote the horizontal and vertical directions at the sample plane, respectively. Then, substituting Eq. (3) into Eq. (1) gives the electric field at the far-field $\tilde{E}$ as:

$$\begin{bmatrix} \tilde{E}_H(u,v) \\ \tilde{E}_V(u,v) \end{bmatrix} \propto \frac{n_m}{\lambda} \tilde{\mathbf{J}} \begin{bmatrix} \cos\alpha \\ \sin\alpha \end{bmatrix} = \frac{n_m}{\lambda}\begin{bmatrix} \tilde{J}_{HH}(u,v) & \tilde{J}_{HV}(u,v) \\ \tilde{J}_{VH}(u,v) & \tilde{J}_{VV}(u,v) \end{bmatrix}\begin{bmatrix} \cos\alpha \\ \sin\alpha \end{bmatrix}, \qquad (4)$$

where $\tilde{J}_{pq}(u,v)$ denotes the 2D spatial Fourier transform of $J_{pq}(x,y)$.

Equations (2)–(4) show how the polarization-sensitive far-field scattering patterns can be retrieved using polarization-sensitive DPM and FTLS. The matrix $\tilde{\mathbf{J}}$ relates between the electric fields at the far plane and the incident field, which can be understood as a representation of a scattering amplitude matrix [29] with respect to the basis vectors of spatial frequencies, $u$ and $v$. Conventionally, the scattering amplitude matrix is represented with respect to the basis vectors of $\theta$ and $\varphi$ directions. For example, $\tilde{J}_{HV}$ implies the complex field scattered into the *horizontal* component from the *vertical* component. It is equivalent to the complex field at the far plane when the polarizer and analyzer are aligned to vertical and horizontal directions, respectively. The polarization-dependent ALS pattern $I_{pq}(u,v)$ can be obtained by taking a modulus square of the Fourier transformed Jones matrix, i.e., $I_{pq}(u,v) = \left|\tilde{J}_{pq}(u,v)\right|^2$.

*2.4 Sample preparation*

To make an isolated LC microdroplet, we added an LC solution (5CB, 328510, Sigma-Aldrich) to the deionized water with sodium dodecyl sulfate (SDS, Fluka). Final volume ratio for each component was 94 (water) : 5 (5CB) : 1 (SDS), which was empirically determined. To fabricate micrometer-sized droplets, the solution was thoroughly agitated using a vortex mixer. Then, 5 mL of the mixed solution was dropped on a coverslip, which was then sandwiched by another coverslip. The boundary between the coverslips was sealed with nail polish to prevent evaporation of the water. The sandwiched sample was heated at 45 °C for 10 min on a heating plate and then cooled at room temperature, so that the LC droplet had a single nematic domain.

Commercial silver NWs, with the average diameter of 60 nm, immersed in isopropanol (739421, Sigma-Aldrich) were prepared by the manufacturer. The NWs were further diluted with the factor of 20 with isopropanol. Then 5 mL of the diluted NW solution was sandwiched between coverslips and sealed by nail polish to prevent evaporation of isopropanol. All measurements were performed at room temperature.

### 2.5 Numerical simulation of spectro-angular light scattering of a silver nanowire

To calculate light scattering intensity of the silver NW, we conducted three-dimensional (3D) finite-difference-time-domain (FDTD) simulation and employed the near-to-far-field transformation based on the transfer matrix method and the reciprocity in electromagnetism [30]. The diameter and the length of NW were set as 60 nm and 4 µm, respectively. The NW was simulated to be immersed in isopropanol ($n = 1.38$) supported on a glass substrate ($n = 1.50$). The domain and grid sizes for the simulation were $1200 \times 5200 \times 1000$ nm$^3$ and 10 nm, respectively. To obtain the scattering intensity distribution at the plane perpendicular to the NW length direction, we calculated the scattered far-field distribution using the total-field/scattered-field method.

## 3. Results

### 3.1 Measurement of light scattering from a liquid crystal droplet

To demonstrate the capability of the pFTLS method, we measured the scattered electric fields of the LC droplet. LC droplets exhibit strong birefringence induced by molecular alignment. Using polarization-sensitive DPM, the spatially resolved Jones matrix of the LC droplet immersed in water was measured (See Method).

The measured spatially resolved Jones matrix and corresponding scattered fields of the LC droplet are shown in Fig. 2. The off-diagonal elements $J_{HV}$ and $J_{VH}$ of the Jones matrix of the LC droplet show fan-like patterns [Fig. 2(a)]. This pattern is consistent with the characteristic birefringence of a radially aligned LC [31]. Insignificant signals are observed at the background of $J_{HV}$ and $J_{VH}$ because there is no birefringence.

From the measured Jones matrix, the scattered complex fields of the LC droplet were calculated using FTLS. The scattered field amplitudes of the LC droplet are shown in Fig. 2(b). Unlike light scattering from a homogeneous sphere, the scattered field amplitudes of the LC droplet show the unique patterns having fourfold rotational symmetry. This symmetry is more clearly visible in the cross-polarizer configurations (*HV* and *VH*). The vanishing lines, indicating extremely weak scattering, are observed along *u*- and *v*-axis in $\tilde{J}_{HV}$ and $\tilde{J}_{VH}$, which are consistent with the previous report [32].

The polarization-dependent ALS intensities along $\theta$ at the four scattering planes ($\varphi = 0°, 45°, 90°,$ and $135°$) are presented in Fig. 2(c). Due to the symmetry for $-\theta$ and $+\theta$ in this scattering geometry, the intensities at $-\theta$ and $+\theta$ were averaged and presented along the positive $\theta$ direction. The scattering angle presented here is the angle in the surrounding medium. The fourfold rotational symmetry of the scattering intensity in cross-polarizer configurations is clearly observed in Fig. 2(c).

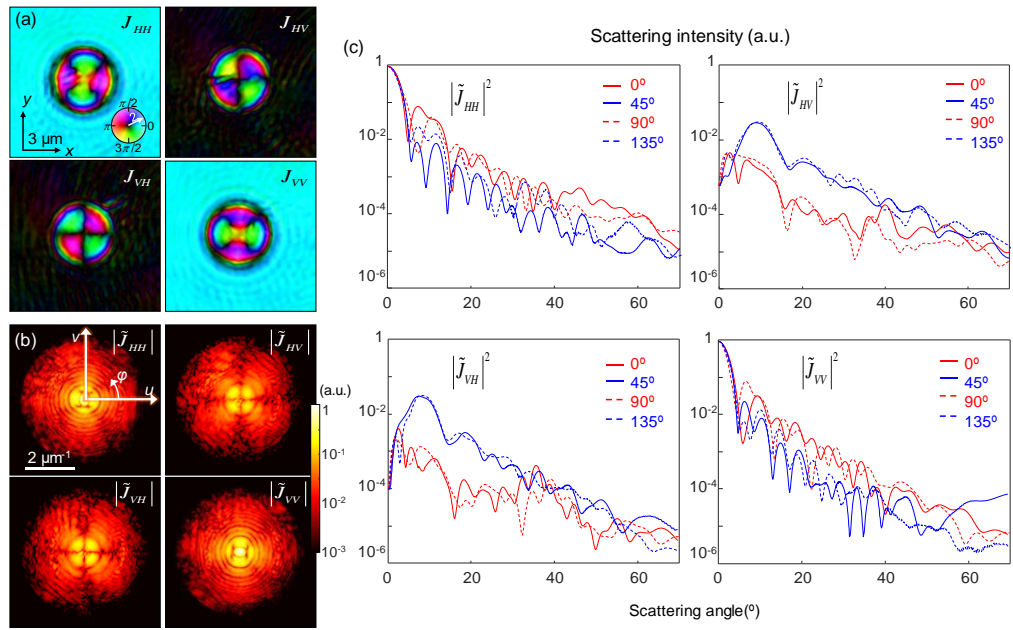

Fig. 2. (a) Measured spatially resolved Jones matrix of the LC droplet. Brightness and color present a modulus and a phase of a complex value, respectively. (b) Polarization-dependent optical fields in the scattering plane. (c) Variation of light scattering intensity with the scattering angle $\theta$ at four representative scattering planes with $\varphi = 0°$, 45°, 90°, and 135°, respectively. Azimuthal angle $\varphi$ is defined to be zero at the positive $x(u)$-axis as depicted in (b).

## 3.2 Measurement of light scattering from a silver nanowire

To demonstrate the capability of the pFTLS method, we also measured the light scattering from the silver NW, which is a morphologically anisotropic sample. The 2D Jones matrix of the vertically aligned NW measured at $\lambda = 530$ nm is presented in Fig. 3(a). Because of the geometric asymmetry of the NW for horizontal and vertical polarized light, $J_{HH}$ and $J_{VV}$ have different complex values that indicate the different responses of the NW to the incident polarization. In contrast to the LC droplet results, however, there is no significant signal in $J_{HV}$ and $J_{VH}$. This indicates that the parallel and perpendicular directions to the long-axis of the silver NW are principal optics axes.

The scattered optical field amplitudes of the NW are presented in Fig. 3(b), which show a line-shaped pattern perpendicular to the direction of the NW. The amplitude along the horizontal plane ($\varphi = 0°$) shows strong signals over large angles, whereas the amplitude along the vertical scattering plane ($\varphi = 90°$) disappears quickly. These scattering patterns are consistent with the extremely thin (60 nm) geometry of the NW.

The light scattering intensities along the scattering angle at the scattering planes perpendicular ($\varphi = 0°$) and parallel ($\varphi = 90°$) to the long axis of the NW are shown in Fig. 3(c). The results present that most of the scattered energy is distributed along the perpendicular direction to the NW. The light scattering intensity in the perpendicular direction ($\varphi = 0°$) is approximately three orders of magnitude stronger than the scattering to the parallel direction ($\varphi = 90°$) in the *HH* and *VV* components. Because the principal optical axes of the NW are parallel to the basis vectors of the matrix, the amount of depolarization energy, i.e., intensity in *HV* and *VH* configuration, is negligible.

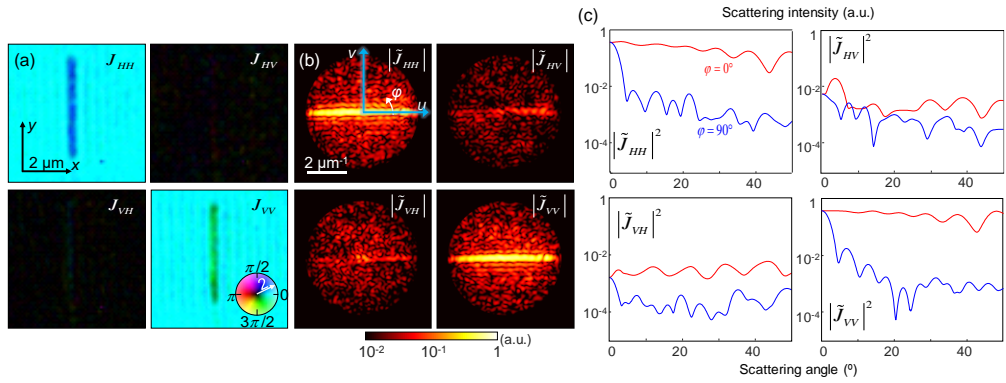

Fig. 3. Measured Jones matrix and scattered field of a silver NW. (a) Spatially resolved Jones matrix of the vertically aligned silver NW. Brightness and color present a modulus and a phase of a complex value, respectively. (b) Amplitudes of the light scattering fields obtained using the pFTLS method. (c) Variation of light scattering intensity profiles with the scattering angle $\theta$ at the perpendicular ($\varphi = 0°$) and parallel ($\varphi = 90°$) scattering planes to the NW.

To further investigate the polarization dependence of the NW scattered light, we emulated the light-scattering intensity of the NW from the obtained $\tilde{\mathbf{J}}$. Because $\tilde{\mathbf{J}}$ fully provides the relationship between the polarization components of the fields, the scattered fields with arbitrary polarization could be inferred from the coherent summation of each $\tilde{J}_{pq}$ with appropriate coefficients. For the incident field with the polarization angle $\alpha$, the light scattering intensity is expressed as follows:

$$I_{Total} = I_H + I_V = \left|\tilde{J}_{HH}\cos\alpha + \tilde{J}_{HV}\sin\alpha\right|^2 + \left|\tilde{J}_{VH}\cos\alpha + \tilde{J}_{VV}\sin\alpha\right|^2, \quad (5)$$

where $I_H$ and $I_V$ represent the horizontally and vertically polarized scattering intensities, respectively. According to Eq. (5), the polarized light scattering intensity toward the direction of $\theta = 30°$ and $\varphi = 0°$ as a function of the incident polarization angle $\alpha$ is shown in Fig. 4. The results present that the maximum total light scattering is achieved when the incident polarization is parallel to the direction of the NW.

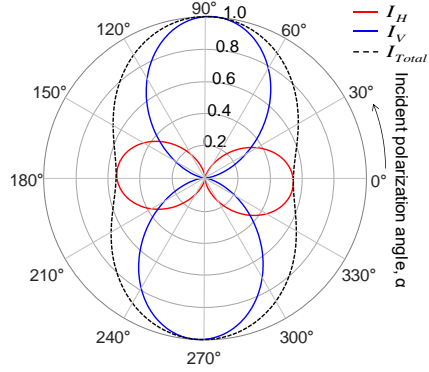

Fig. 4. Normalized light scattering intensity from the vertically aligned NW shown in Fig. 3 at the direction of $\theta = 30°$ and $\varphi = 0°$. The maximum scattering intensity occurs when the incident polarization is parallel to the long axis of the NW. The red and blue solid lines represent the intensity of the horizontally ($I_H$) and vertically ($I_V$) polarized scattered light, respectively. The dashed line represents the sum of intensity for both polarizations ($I_{Total}$).

*3.3 Spectroscopic extension of the pFTLS*

The utilization of the pFTLS is not limited to monochromatic light. We demonstrate that the present method can be extended to spectroscopic 2D polarized light scattering with wavelength-scanning illumination. For verification, we measured polarization-dependent scattering light intensity for four individual silver NWs at 50 wavelengths in the range of 470–715 nm as shown in Fig. 5(a).

The scattering intensity shows distinct spectro-angular distribution depending on the incident polarization. For the horizontal polarization illumination (i.e., perpendicular to the NW), the scattering intensity decreases as the wavelength and scattering angle increase. In the case of vertical polarization illumination (i.e., parallel to the NW), by contrast, the scattering intensity is nearly constant along the scattering angle and wavelength within detected ranges. The spectral intensity in $I_{VV}$ can be explained by the NW length that is considerably longer than the wavelengths. Regardless the illumination wavelengths we used, the NW is always considerably longer compared to the wavelength of the light. The difference in the angular dependency between $I_{HH}$ and $I_{VV}$ could be understood by the different orientation of an induced dipole in the NW. Because the dipole does not radiate energy along the oscillating direction, the scattering intensity at the horizontal plane ($\varphi = 0°$) should gradually decrease along the scattering angle and becomes zero at $\theta = 90°$ for the horizontal polarization case.

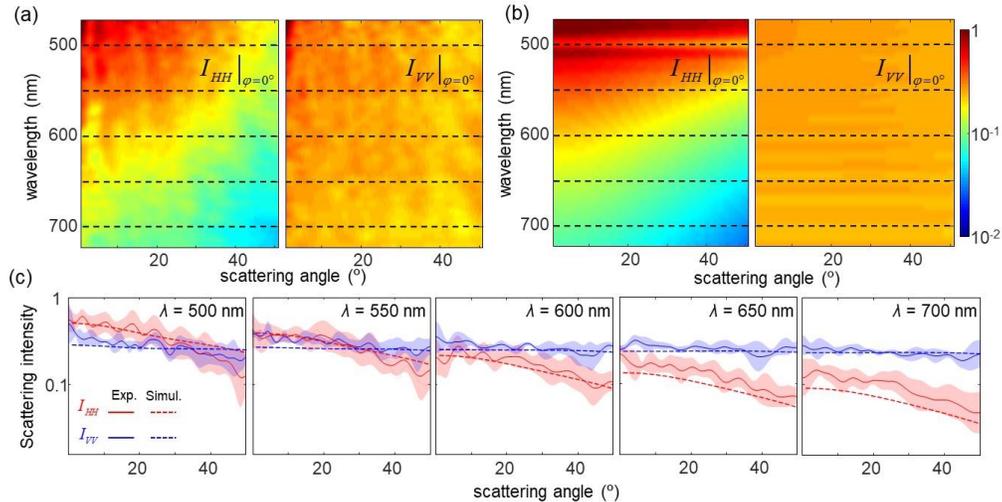

Fig. 5. Spectroscopic pFTLS signals of the NWs. (a) Measured spectro-angular light scattering intensity averaged from four NWs. (b) Numerically calculated results using the FDTD method. (c) Comparison between the experimental and numerical results at five representative wavelengths denoted by dashed lines in (a) and (b). The solid and dashed lines represent the experiments and simulations, respectively. The shaded areas indicate the standard deviations of the experiment results from four NWs.

The numerical simulation results using the FDTD method [Fig. 5(b)] exhibit a similar spectro-angular distribution of the scattering intensity. The detailed comparisons at five representative wavelengths ($\lambda$ = 500, 550, 600, 650, 700 nm) shown in Fig. 5(c) present good agreements between experiments and simulations, except for slightly uneven experimental noises.

## 4. Discussion and conclusion

We presented the pFTLS technique that enables the measurement of polarization-dependent ALS. Utilizing polarization-sensitive DPM and FTLS, the polarization-dependent ALS fields of individual microscopic samples were precisely measured without angular scanning. The capability of the pFTLS was verified by measuring the LC droplet and the silver NW. The measured scattered fields of both samples show distinct angular scattering patterns reflecting the characteristic birefringence of the samples. In addition, we demonstrated the spectroscopic pFTLS on the individual silver NWs using the wavelength-scanning illumination. In summary, this method enables access to the information on light scattering based on the scattering angle, polarization, and wavelength over broad angular and spectral ranges.

Compared to the conventional approaches, the present approach has several significant advantages. First, the pFTLS allows measuring polarization-sensitive ALS signals from individual microscopic objects with weak scattering cross-sections. Because it only requires the four measurements of holograms to retrieve the polarization-sensitive 2D ALS signals, high-speed and high-throughput measurements are available. It may enable the measurement of dynamic polarization-dependent ALS signals, which can be exploited to study biological objects exhibiting birefringences such as red blood cells from sickle cell disease [24, 33, 34] or crystals of monosodium urate in gout patients [35, 36].

From the technical point of view, the pFTLS method can be readily applied to the data obtained with polarization-sensitive QPI techniques [19, 37-46]. From the measured polarization-sensitive optical field images, polarization-sensitive ALS signals can also be extracted.

We expect the wealth of information would bring unprecedented opportunities for theoretical studies and practical applications of the polarization light scattering, for example, from the metallic nano- and micro-structures [47], colloidal particles [48], and biological tissues and cells [49, 50].

## Acknowledgements

This work was supported by KAIST, BK21+ program, Tomocube, and National Research Foundation of Korea (2015R1A3A2066550, 2014M3C1A3052567, 2014K1A3A1A09063027, 2017R1A2B2009117). The authors thank Mr. Yun-Seok Choi for useful discussion.